\documentclass[9pt,twocolumn,twoside]{optica}
\setboolean{shortarticle}{false}
\setboolean{minireview}{false}

\usepackage{epsfig}
\usepackage{dcolumn}
\usepackage{bm}
\usepackage{xcolor}

\newcommand{\Om}{\Omega}
\newcommand{\om}{\omega}
\newcommand{\Omeff}{\Omega_{\rm eff}}
\newcommand{\OmAC}{\Omega_{\rm AC}}

\newcommand{\Omsgl}{R_{\rm sc}}
\newcommand{\Isat}{I_{\rm sat}}
\newcommand{\omHFS}{\omega_{\rm hs}}

\title{Multiaxis atom interferometry with a single diode laser and a pyramidal magneto-optical trap}

\author[1,*]{Xuejian Wu}
\author[1,2]{Fei Zi}
\author[1]{Jordan Dudley}
\author[1]{Ryan J. Bilotta}
\author[1]{Philip Canoza}
\author[1,3]{Holger M\"uller}

\affil[1]{Department of Physics, University of California, Berkeley, CA 94720, USA}
\affil[2]{Institute of Optics, Department of Physics, Zhejiang University, Hangzhou 310058, China}
\affil[3]{Molecular Biophysics and Integrated Bioimaging, Lawrence Berkeley National Laboratory, Berkeley, CA 94720, USA}

\affil[*]{Corresponding author: xuejianw@berkeley.edu}

\ociscodes{(020.1335) Atom optics; (120.3940) Metrology; (020.3320) Laser cooling; (120.3180) Interferometry.}

\begin{abstract}
Atom interferometry has become one of the most powerful technologies for precision measurements. In order to develop simple, precise and versatile atom interferometers for inertial sensing, we demonstrate an atom interferometer measuring acceleration, rotation, and inclination by pointing Raman beams toward individual faces of a pyramidal mirror. Only a single diode laser is used for all functions, including atom trapping, interferometry, and detection. Efficient Doppler-sensitive Raman transitions are achieved without velocity selecting the atom sample, and with zero differential AC Stark shift between the cesium hyperfine ground states, increasing signal-to-noise and suppressing systematic effects. We measure gravity along two axes (vertical and 45$^\circ$ to the vertical), rotation, and inclination with sensitivities of 6\,$\mu$m/s$^2/\sqrt{\rm Hz}$, 300\,$\mu$rad/s/$\sqrt{\rm Hz}$, and 4\,$\mu$rad/$\sqrt{\rm Hz}$, respectively. This work paves the way toward deployable multiaxis atom interferometers for geodesy, geology, or inertial navigation.
\end{abstract}

\setboolean{displaycopyright}{true}

\begin{document}

\maketitle

\section{Introduction}

Light-pulse atom interferometers (AIs) \cite{Cronin} use the recoil momentum from photon-atom interactions to coherently split and recombine matter waves. They have been used for measuring gravity \cite{Peters,GravIso,Hu}, the gravity gradient \cite{McGuirk,Rosi,Asenbaum}, rotation \cite{Gustavson,stockton,Dutta}, fundamental constants \cite{Fixler,Rosi2014,Lan,Bouchendira,Parker}, and for testing fundamental laws of physics \cite{Zhou,Hartwig,Duan,Hamilton,Jaffe,Kovachy,Yu,Graham,Harms}. Since the laser wavelength defines the photon momentum with high precision, AIs are accurate. Thus they are ideal candidates for inertial sensing or navigation. For this purpose, AIs need to be simple, reliable, and sensitive to multiple axes of acceleration and rotation. Even transportable single-axis AIs, however, require several lasers and laser amplifiers for atom trapping, interferometry, and detection \cite{Freier,Fang,Geiger,Barrett}. So far, the only AI with six-axis sensing 
utilized two parabolically launched atom clouds and a complex combination of separate interferometry setups \cite{canual}. It achieved a sensitivity of 22\,$\mu$rad/s/$\sqrt{\rm Hz}$ and 16\,$\mu$m/s$^2/\sqrt{\rm Hz}$ for rotation and acceleration, respectively. 
Two-axis of rotations and one-axis of acceleration have also been demonstrated in an atomic fountain interferometer with atomic point sources and spatially resolved detection \cite{Dickerson}. An atomic sensor using Bose-Einstein condensates has simultaneously measured gravity and magnetic field gradients \cite{Hardman}. A dual-axis accelerometer and gyroscope atom interferometer has been built by launching and recapturing two cold ensembles toward each other \cite{Rakholia}. These examples illustrate that multiaxis AIs are more complex than single-axis ones. Additionally, other advances towards field operations, such as cold atom pyramidal gravimeters \cite{Bodart}, atom interferometers with short integration time \cite{Butts}, atom interferometers with optical lattice \cite{Andia}, and atom interferometry in an optical cavity \cite{CavityAI} or a warm vapor \cite{Biedermann}, and atom-chip gravimeters \cite{Abend} have been demonstrated as well. However, multiaxis operation and simplicity have yet to come together in AIs.

Generally, the laser system contributes the most complexity. Magneto-optical traps (MOTs) require six orthogonal beams and matter-wave splitters need relatively high laser intensity and low phase noise. Besides, specific laser frequencies are demanded for different procedures. In order to construct simple and reliable laser systems, fiber lasers \cite{fiberlaser} and integrated diode lasers \cite{diodelaser} are developed. However, to our knowledge, AIs have never been operated based on a single diode laser without optical amplifiers. Laser systems with a single diode laser and pulsed modulators can avoid frequency-locking or phase-locking between different lasers and thus improve robustness. Without optical amplifiers, laser systems will also gain simplicity and power efficiency. 

Here, we demonstrate a multiaxis AI based on a single diode laser and a pyramidal MOT. 
The pyramidal geometry requires only a single laser beam to trap atoms and form a vertical atom interferometer. Additional beams, orthogonal to the pyramidal faces, allow for a total of five AIs along different axes. Using the Mach-Zehnder geometry and the butterfly geometry 
allows for measuring acceleration and rotation separately. A single diode laser serves the multiaxis AI to maintain simplicity. With efficient two-photon Raman transitions and zero differential AC Stark shift, high-contrast fringes have been achieved using a $\mu$K-sample without velocity selection. As a demonstration, we achieve a sensitivity of 6\,$\mu$m/s$^2/\sqrt{\rm Hz}$, 300\,$\mu$rad/s/$\sqrt{\rm Hz}$, and 4\,$\mu$rad/$\sqrt{\rm Hz}$ for acceleration, rotation, and inclination, respectively, limited by vibrational noise. This work offers a path towards building simple, precise and multiaxis AIs.


\section{Multiaxis atom interferometry}

\begin{figure}
\centering
\epsfig{file=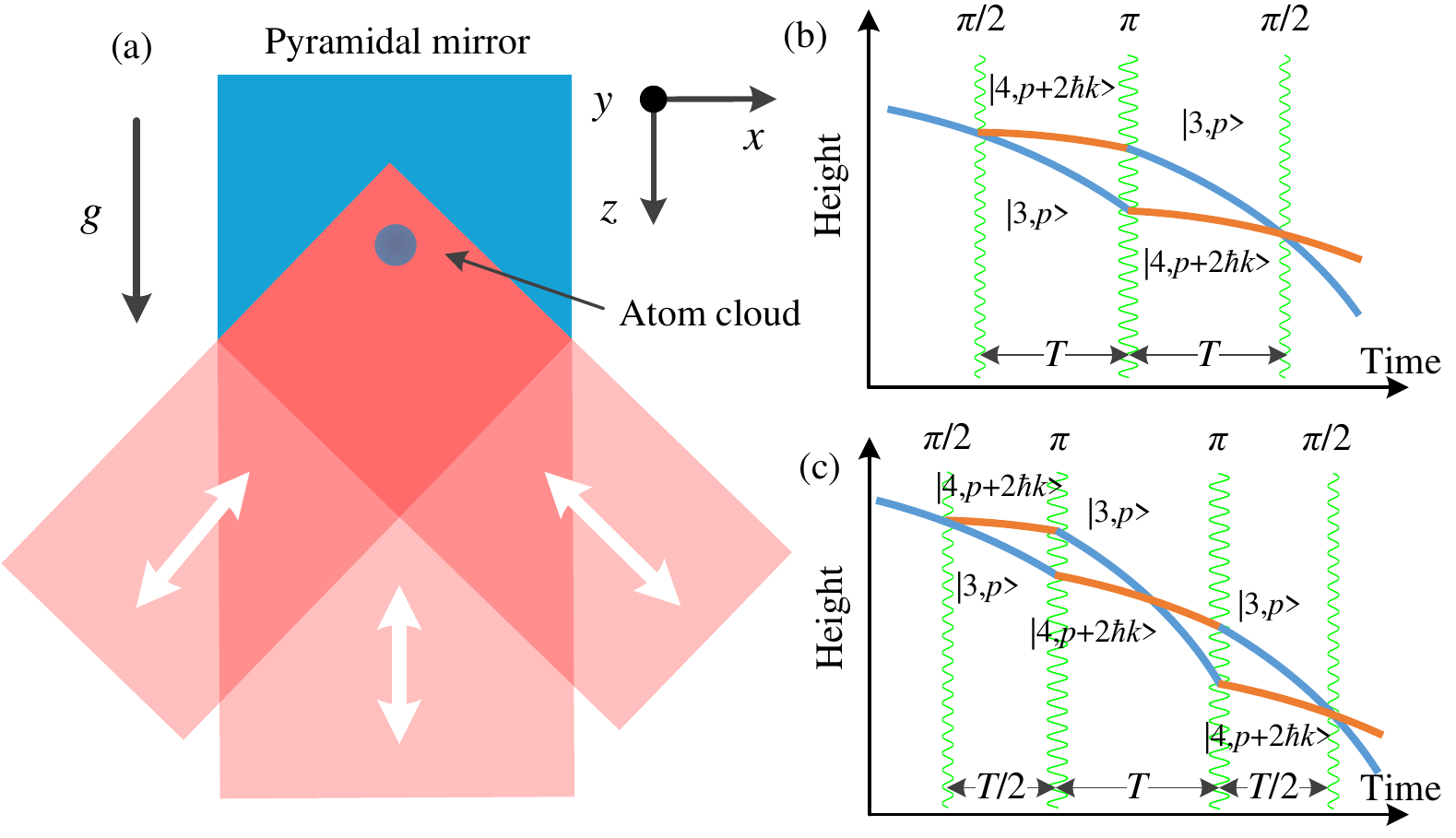,width=\linewidth}
\caption{\label{Pyramid} (a) Multiaxis AIs in a pyramid. A cold atom cloud is trapped inside a pyramidal mirror with a top angle of 90$^\circ$. Five pairs of the retro-reflected Raman beams are formed, one along the vertical axis and the other four perpendicular to the pyramidal faces. The two angled Raman pairs in the $yz$ plane are not shown. The angled Raman beams are approximately 45$^\circ$ to the gravity axis. (b) Space-time trajectories of atoms in the Mach-Zehnder geometry and (c) in the butterfly geometry. A matter wave (blue and orange curves) is coherently split, redirected and combined by momentum transfer from laser pulses (green waves).}
\end{figure}

Figure \ref{Pyramid}(a) shows the principle of the multiaxis atom interferometry in a pyramid. The pyramid consists of four orthogonal reflection faces. A MOT is created inside the pyramid by irradiating one laser beam vertically toward the entire pyramid, where six orthogonal trapping beams can be generated by the reflections \cite{Lee,Pollock}. Utilizing the incidence and its reflections from either the whole pyramid or individual pyramidal faces as matter-wave splitters, we can build  one vertical AI as well as four angled AIs along different axes. 

The matter-wave splitter of our AI is based on Doppler-sensitive two-photon Raman transitions  between the $F=3$ and $F=4$ hyperfine ground states of cesium atoms \cite{KasevichChu}. An atom, initially in the state $|F=3, p=0\rangle$, is transferred to a state $|F=4, p=2\hbar k\rangle$, where $\hbar k$ is the photon momentum. To make a beam splitter, a $\pi/2$ pulse places the atom in a superposition of the two states. A mirror is formed by a $\pi$-pulse, which has a 100\% probability of changing the state.


As shown in Fig. \ref{Pyramid}(b), Mach-Zehnder interferometry is performed by a $\pi/2 -\pi-\pi/2$ pulse sequence and is used for measuring acceleration. We assume $\vec{a}\cdot\vec{\Omega} T_0 $ and $\vec{\Omega} \cdot \vec{v}_0$ are negligible compared to the acceleration $\vec a$ and the gravity $\vec g$, where $\vec{\Omega}$ is the rotation, $T_0$ is the sequence time, and $\vec{v}_0$ is the initial velocity of the atom cloud at the first laser pulse. This condition is fulfilled, e.g., in stationary operation or aboard moving vehicles. The phase shift \cite{canual} caused by $\vec a$ and $\vec g$ is expressed as 
\begin{equation}
\phi_a = \vec k\cdot (\vec a + \vec g )T^2,
\end{equation}
where $\vec k$ is the effective wave vector of two counterpropagating photons and $T$ is the pulse separation time. A vertical interferometer (addressed by the vertical laser beam and its reflection from the pyramid) and at least two angled ones (formed by beams aimed directly at a pyramid face) allows us to measure the full acceleration vector $\vec a=(a_x, a_y, a_z)$.  

To measure rotation independent of acceleration, we use the butterfly geometry with a $\pi/2-\pi-\pi-\pi/2$ pulse sequence, as shown in Fig. \ref{Pyramid}(c). The rotation-induced phase shift \cite{canual,stockton} is 
\begin{equation}
\phi_\Om =\frac 12 \vec k \cdot [(\vec a +\vec g)\times \vec \Om]T^3.
\end{equation}
Since $\vec g$ is along the $z$ axis, two components of rotation ($\Om_x, \Om_y$) can be measured using laser beams pointing at two different pyramid faces. Additionally, measurement of $\Om_z$ can be achieved by applying appropriate acceleration in the $xy$ plane to the interferometer. This allows us to measure the full rotation vector $\vec \Om=(\Om_x, \Om_y, \Om_z)$.  
 
\section{Single-diode atom interferometry}

\subsection{Single-diode laser system}

\begin{figure}
\epsfig{file=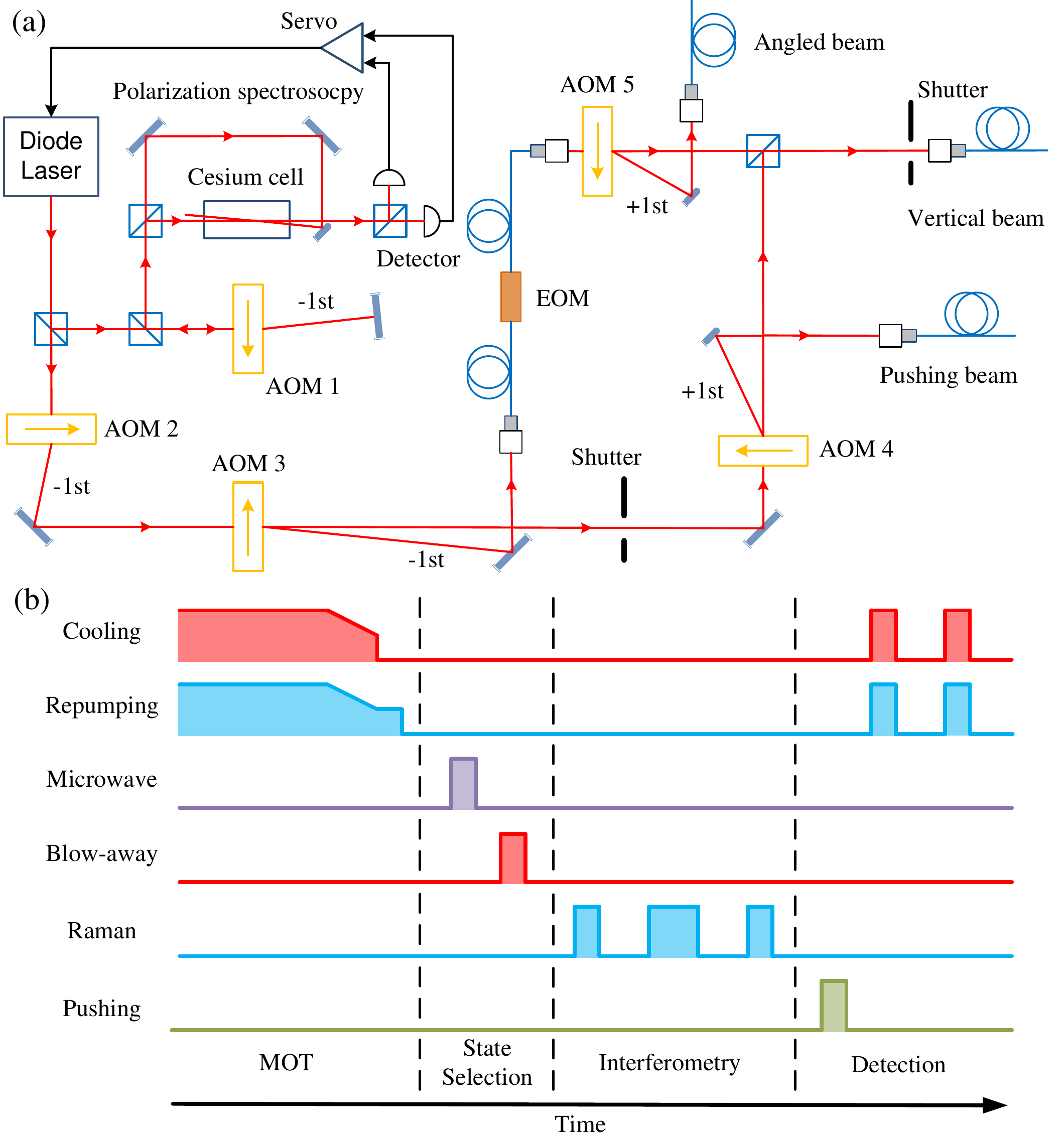,width=0.45\textwidth}
\caption{\label{Laser} (a) Laser system. AOM, acousto-optic modulator; EOM, fiber-based electro-optical modulator. A distributed Bragg reflector diode laser is frequency stabilized by polarization spectroscopy. The frequency detuning of the laser is controlled by AOM 1. AOM 2 works as a fast switch and AOM 3 controls the laser intensity to the EOM. The vertical beam is toward the entire pyramid, and the angled beams are toward individual pyramidal faces. The pushing beam is for spatially separating atoms in the two ground states. (b) Timing sequences of single-laser atom interferometry. Both the cooling beam and the blow-way beam are generated by the undeflected laser of AOM 3. Both the repumping beam and the Raman beam are generated by the EOM.}
\end{figure}

Atom interferometers consist of three procedures: atom cloud preparation, interferometry, and population detection. Only one diode laser is used for all the functions. The laser system is shown in Fig. \ref{Laser}(a). All laser radiation originates from a 240-mW distributed Bragg reflector diode laser (Photodigm, PH852DBR240T8). A sample of its power is sent to Doppler-free polarization spectroscopy, frequency stabilizing (``locking") the laser to the cesium $F=4 \rightarrow F'=4/5$ $D_2$ crossover transition at 852 nm. An acousto-optical modulator (AOM 1) shifts the sample so that the light reaching the atoms can be at the MOT frequency (about 10\,MHz red from the $4-5$ transition) or the detection frequency (resonant with $4-5$). Adding an offset voltage at the servo input jumps the lock point to the $F=4\rightarrow F'=4$ transition and generates the large detuning necessary for polarization gradient cooling. 

The timing sequences of our atom interferometry is shown in Fig. \ref{Laser}(b). The MOT cooling light is the undeflected beam after AOM 3; a repumping frequency is generated by sending a sample of the laser through a fiber electro-optical modulator (EOM). To avoid instability resulting from interference with the MOT light, the EOM is driven such that the carrier frequency is nulled. A liquid crystal retarder is placed after the fiber to convert the linear polarization to the circular polarization, so that counterpropagating $\sigma^+/\sigma^-$ polarization pairs are formed inside the pyramid. Before interferometry, a microwave pulse followed by a blow-away laser pulse (resonant with $4-5$) selects atoms into the magnetically insensitive state.

The Raman frequency pairs for the interferometer are formed by the carrier and the first-order sidebands from the EOM. The Raman pulses have high intensity in the fiber EOM, but their pulse duration is too short to cause photo-refractive damage to the crystal. To minimize phase noise, the EOM is driven by a phase-locked dielectric resonator oscillator. The pyramidal geometry enables lin $\perp$ lin polarization in the vertical AI, and $\sigma^+/\sigma^+$ or $\sigma^-/\sigma^-$ polarization in the angled AIs. The vertical laser to the pyramid is blocked by a shutter when the angled AIs are operated.

For detecting the $F=3$ and $F=4$ populations at the interferometer output ports, a pushing beam, slightly red-detuned to the $F=4 \rightarrow F'=5$ transition, horizontally separates atoms in the two hyperfine ground states \cite{Biedermann2009}. Both the cooling beam and repumping beam are then used for fluorescence imaging. A camera images both populations simultaneously, which makes the interference fringe immune to the fluctuation of atom number and imaging laser power.

\subsection{Zero differential AC Stark shift with small detuning}

\begin{figure}
\epsfig{file=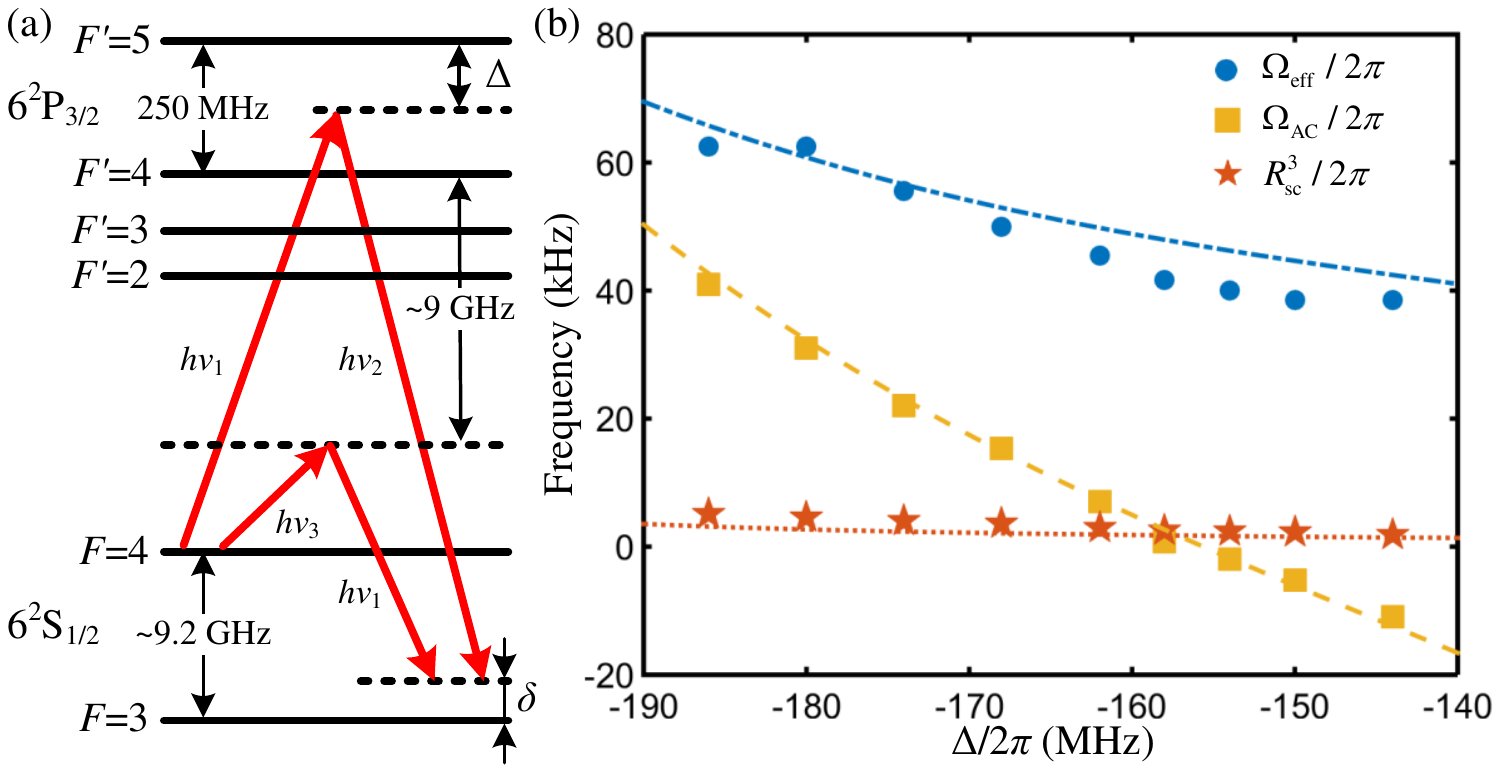,width=0.45\textwidth}
\caption{\label{Levels} (a) Energy level scheme of cesium $D_2$ line. 
The Raman pairs are formed by the carrier $\nu_1$ and the first-order sidebands $\nu_2$, $\nu_3$ from the EOM. (b) Rabi frequency $\Omeff$, AC Stark shift $\OmAC$ and single photon scattering $\Omsgl$ as a function of single photon detuning $\Delta$. The Rabi frequency and AC Stark shift are measured by driving two-photon transitions. The measurement uses atoms inside the pyramid, where they see reflections from the four pyramidal faces. This increases $\OmAC$ and $\Omsgl$ but not $\Omeff$. The scattering rate $\Omsgl^3$ is measured as the number of atoms that are transferred from $F=3$ to $F=4$ when the Raman detuning is off the two-photon resonance. Each point is a single experimental shot and the curves are the theory predictions.}
\end{figure}

In order to drive Raman transitions with modest laser intensity, the performance with a small single photon detuning is investigated. Figure \ref{Levels}(a) shows the energy levels and laser frequencies involved in the Raman transitions. The transitions must satisfy several requirements. Rapid Raman transitions ($\pi$-pulse time 10-20\,$\mu$s) are needed in order to address all atoms from the thermal velocity distribution of the MOT efficiently, but it requires high laser intensity and/or small single-photon detuning $\Delta$. For high accuracy and fringe contrast, the AC Stark shift of the $F=3$ and $F=4$ states needs to be equal, so that it cancels out of the interferometer phase \cite{Peters}. We calculate the effective two-photon Rabi frequency $\Omeff$, the differential AC Stark shift $\OmAC$, and the single-photon scattering rate $\Omsgl$. To do so, we define $A_n=\sqrt{I/\Isat}\Gamma J_n(\beta)/\sqrt{2}$ to describe the amplitude of each EOM sideband, where $I$ is the total laser intensity, $\Isat$ is the saturation intensity, $\Gamma$ is the linewidth, $J_n$ is the Bessel function of order $n$ and $\beta$ is the modulation index of the EOM. With this, we have
\begin{eqnarray}
\Omeff&=& \sum_{F'=2}^5\sum_{n=-\infty}^\infty \frac{M_{3,0}^{F',-} A_n M_{4,0}^{F',+} A_{n+1}}{2\Delta_3},\\
\OmAC&=&\ \sum_{F',n}\left(\frac{|M_{3,0}^{F',-}A_n|^2}{4\Delta_3}-\frac{|M_{4,0}^{F',+}A_n|^2}{4\Delta_4}\right),
\end{eqnarray}
where $M_{F, m_F}^{F',\pm} = \langle F, m_F|F', m_F\pm 1 \rangle$ are the cesium $D_2$ dipole matrix elements for $\sigma^\pm$ transitions, expressed as multiples of $\langle J = 1/2||er||J' = 3/2\rangle$ \cite{Steck}, $\Delta_F=n\omHFS+\om_F^{F'}+\Delta$ is the effective detuning of $F=$ 3 or 4, $-\hbar \om_4^{F'}$ is the energy of the $|F'\rangle$ excited state relative to the $|F'=5\rangle$ state, and $\om_3^{F'}=\om_4^{F'}-\omHFS$. The ground state hyperfine splitting $\omHFS$ is also the EOM driving frequency, and $\Delta$ is the detuning of the carrier relative to the $F=4\rightarrow F'=5$ transition. The scattering rates for atoms starting in $F=3$ or $F=4$ are
\begin{equation}
\Omsgl^F=\sum_{F',n}\frac{\Gamma(M_{F,0}^{F',-}A_n)^2}{\Gamma^2+2(M_{F,0}^{F',-}A_n)^2+4(\Delta_F)^2}. 
\end{equation}
From these, the scattering probability for an entire interferometer can be calculated. 

A suitable detuning $\Delta/(2\pi) \simeq -160\,$MHz (slightly dependent on $\beta$), can be found for which $\OmAC$ vanishes and both $\Omeff$, $\Omsgl$ are acceptable. This detuning can easily be reached with an AOM. In this case, the theoretical limit on the contrast of Mach-Zehnder fringes from single-photon scattering is approximately $40\%$.

Figure \ref{Levels}(b) shows a comparison of theory to experiment for the two-photon Rabi frequency, differential AC Stark shift, and single-photon scattering as function of detuning. If we modulate the fiber EOM with an index of about 1, the differential AC Stark shift is zeroed at a red single photon detuning of 158\,MHz. The zero differential AC Stark shift is verified with 30-Hz accuracy by measuring $\omHFS$ with optical Ramsey interferometry. For the Doppler-sensitive Raman transition, the width of a $\pi$-pulse is as short as 12\,$\mu$s. Since Doppler-sensitive Raman transitions can only transfer atoms distributed within a certain velocity bandwidth, the efficiency of Raman transitions can be increased by use of a faster Rabi flopping frequency. Without velocity selection, a $\pi$ pulse transfers as many as 60\% of all atoms.

\section{Experiment and Results}

\subsection{Experiment}

The pyramid is in a glass cube of $25.4\times 25.4\times 25.4$\, mm$^3$, dielectrically coated for equal phase shift at two orthogonal polarizations at 45$^\circ$. The vertical AI (and MOT and detection) beam has a waist of 15 mm ($1/e^2$ radius) and a power of $\sim 60\,$mW before the pyramid. The diagonal beams have waists of 12 mm ($1/e^2$ radius) and power of $\sim 40\,$mW.

We capture approximately 5 million atoms from background cesium vapor in 1\,s. With increased laser detuning of -160\,MHz and decreased ($\sim 1/4$) laser power, polarization gradients cool the atoms to about 2\,$\mu$K in 5\,ms. The cooling beam is turned off 1\,ms before the repumping beam to ensure that all the atoms stay in $F=4$. As the atoms freely fall, a microwave $\pi$-pulse (100\,$\mu$s and +20 dBm) followed by a blow-way laser pulse selects $\sim 5\times 10^5$ atoms from $|F=4, m_F=0\rangle$ into $|F=3, m_F=0\rangle$ with a bias field of 500\,mG. During interferometry, the magnetic quantization axis is aligned with the direction of the Raman pairs in order to enhance the Raman transition between $|F=3, m_F=0\rangle$ and $|F=4, m_F=0\rangle$.

As a proof of multiaxis atom interferometry, we demonstrate three AIs on a passive vibration isolation platform (minusK, 150BM-1) placed on an optical table. One AI uses the vertical Raman pair, and the other two use angled Raman pairs. The vertical AI is operated with $\pi/2 -\pi-\pi/2$ sequences for gravity measurement. The angled AIs are operated with $\pi/2 -\pi-\pi/2$ sequences and $\pi/2-\pi-\pi-\pi/2$ sequences to measure the projected gravity with an angle of approximately 45$^\circ$ and the earth's rotation rate.

\subsection{Acceleration measurement}

\begin{figure}
\epsfig{file=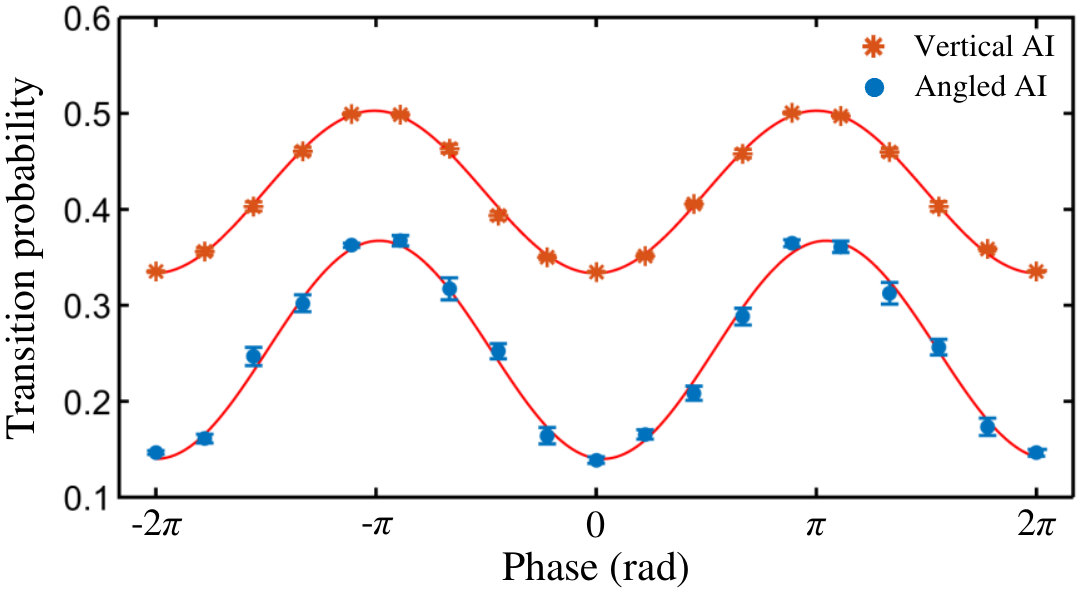,width=0.45\textwidth}
\caption{\label{acceleration} Acceleration-sensitive fringes of Mach-Zehnder AIs. The vertical AI has a sequence time of $2T=80\,$ms. The angled AIs have a sequence time of $2T=40\,$ms. Each point is the average by 10 experimental shots. The curves are sinusoidal fits.}
\end{figure}

Figure \ref{acceleration} shows fringes measured in Mach-Zehnder geometry. 
In the vertical AI, the frequency difference between the Raman frequency pair is linearly ramped at approximately 23\,MHz/s to compensate the time-varying Doppler shift of the free-falling atoms. The ramp rate of the angled AI, at about 45$^\circ$ to the vertical, is about 16\,MHz/s, which is a factor of $\sqrt{2}$ smaller than that of the vertical one. While scanning the phase of the last interferometer pulse, the fringes are obtained by counting the atom populations in the two hyperfine ground states. In particular, the pulse sequence time of the vertical AI is constrained by the geometry of the apparatus and that of the angled AIs is limited by the Raman beam waist. The vertical AI has a fringe constrast of 18\% with a sequence time of 80\,ms. The angled AIs have a fringe constrast of 22\% with a sequence time of 40\,ms. As the sequence time is as short as 2\,ms, the fringe contrast is improved to 30\%. The decreasing contrast with longer sequence time can be explained by inhomogeneous Rabi flopping of the three Raman pulses.  

The top angle of pyramid is 90$^\circ$$\pm$1$^\prime$, which leads to a systematic error of 42 parts per billion for the vertical AI. For the angled AI, the accuracy of the top angle produces a negligible error in the alignment. However, the error of the projection angle results in systematic errors to the accelerations along $x$ or $y$ axis. Given an angle error of 1$^\prime$ at 45$^\circ$, the systematic error would be about 2 mm/s$^2$.

\begin{figure}[!t]
\epsfig{file=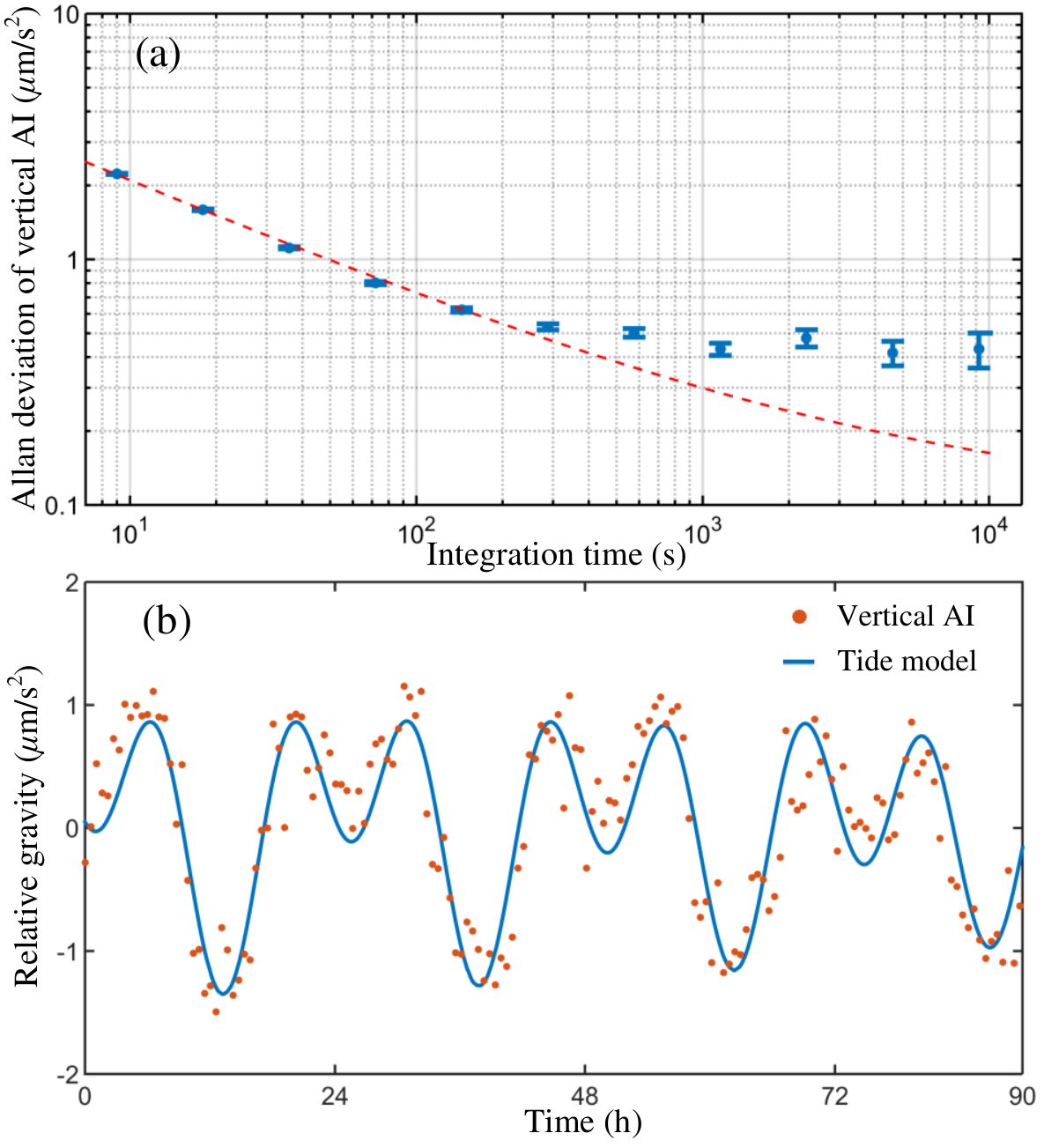,width=0.45\textwidth}
\caption{\label{tide} (a) Allan deviation of the vertical AI corrected from the Earth's tides. The dashed line shows the 1/$\sqrt{\rm Hz}$ scaling. (b) Tide gravity variation. It was measured from August 7, 2017 to August 10, 2017 by the vertical AI, compared with a tidal model. Each data point is averaged for about 30 min.}
\end{figure}

Figure \ref{tide}(a) shows the Allan deviation of the gravity measurement by the vertical AI, with the tide variation subtracted from the model. The sensitivity is 6\,$\mu$m/s$^2/\sqrt{\rm Hz}$. Measuring gravity continuously during 4 days, the tide variation has been  observed, as shown in Fig. \ref{tide}(b). The systematic error of the pyramidal top angle is negligible compared to the current sensitivity. Using a commercial tilt sensor (Jewell Instruments, 756-1326) to monitor the alignment between the Raman beam and the gravity axis, the long-term drift of the platform is calibrated. The sensitivity of the vertical AI is competitive to the state-of-the-art compact AIs \cite{Geiger,Barrett,canual,Rakholia,Bodart,Butts,Andia,CavityAI,Biedermann}.

The sensitivity can be further improved with a faster cycling rate, better vibration isolation, and longer sequence time. Our current cycling time is limited by the computer control software. Once this is overcome, the cycling rate can further be increased by shortening the MOT loading time. Additionally, the vibrational noise can be decreased by better vibration isolation, such as an active feedback system \cite{ZhouM}. Finally, since the sensitivity scales with $T^2$, it can be improved by longer sequence time. For example, a freely-falling time of $\sim$ 300 ms, corresponding to a length of $\sim$ 0.5 m, can improve the current sensitivity by one order of magnitude. 





\subsection{Inclination measurement}

The Mach-Zehnder AIs along the diagonal axes measure the projected gravity with an angle of approximately 45$^\circ$, which makes them sensitive to the angle variation between the Raman beam and the gravity axis. As the fiber ports of the Raman beams are also put on the vibration isolation platform, the angled AIs work as atomic inclinometers.

\begin{figure}
\epsfig{file=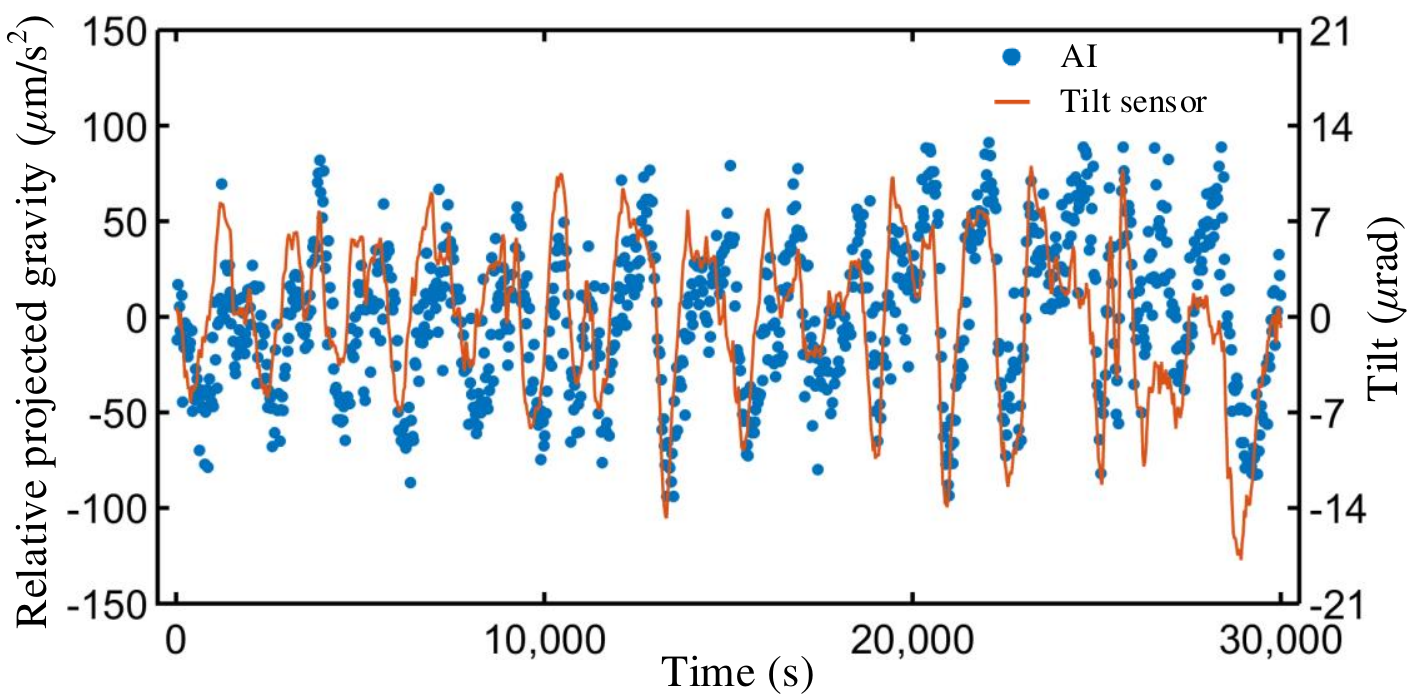,width=0.45\textwidth}
\caption{\label{tilt} Relative projected gravity measured by one of the angled AIs. Each data point is from one fringe, which consists of 11 shots. 
The orange curve shows the tilt measured by a commercial sensor.}
\end{figure}

For long-term measurement, we observe that the fluctuation of the projected gravity is correlated with the tilt sensor, as shown in Fig. \ref{tilt}. 
This fluctuation is the drift of the platform, which has a period of half an hour. It also indicates that the sensitivity of the vertical AI is limited by vibration noise. The sensitivity of the angled AI is about 25\,$\mu$m/s$^2/\sqrt{\rm Hz}$. According to the projection angle, gravity variation of about 150\,$\mu$m/s$^2$ corresponds to the tilt of 21\,$\mu$rad. The sensitivity of our atomic inclinometer is 4\,$\mu$rad/$\sqrt{\rm Hz}$, compared to 800\,$\mu$rad/$\sqrt{\rm Hz}$ in previous work \cite{Ahlers}.

\subsection{Rotation measurement}

\begin{figure}[!t]
\epsfig{file=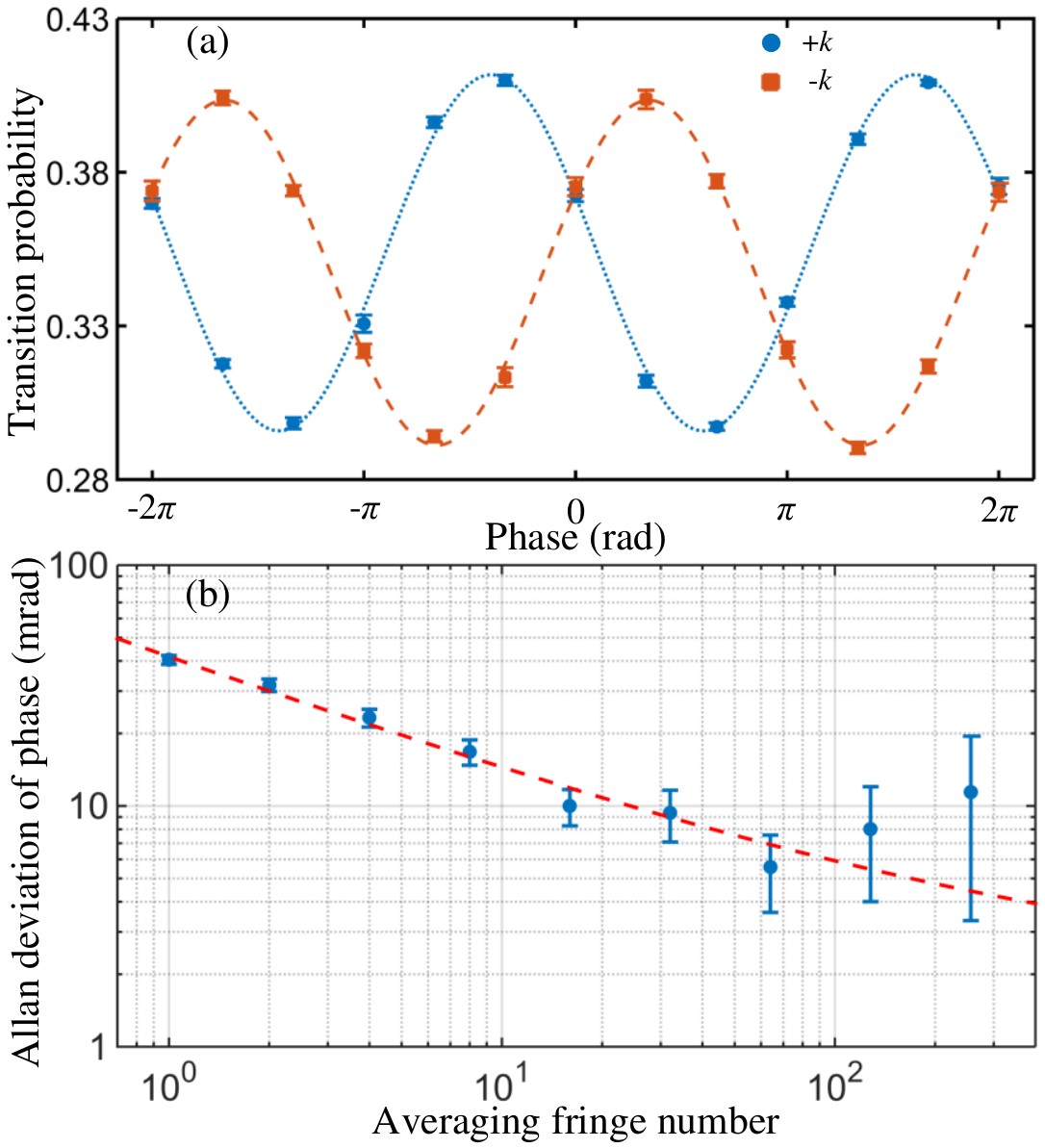,width=0.45\textwidth}
\caption{\label{rotation} (a) Rotation-sensitive fringes of a butterfly AI along the angled axis. The two fringes are obtained with opposite effective wave vectors. The sequence time is $2T=40\,$ms. Each point is the average by 10 experimental shots. The curves are sinusoidal fits. (b) Allan deviation of the gyroscope sensitivity. The dashed line shows the 1/$\sqrt{N}$ scaling, where $N$ is the averaging fringe number. }
\end{figure}

Figure \ref{rotation}(a) shows interference fringes of a symmetric butterfly interferometer (i.e., with pulse separation times $T/2$, $T$, $T/2$) for rotation measurements. With a sequence time of 40\,ms, we achieve a fringe contrast of 11\%. The symmetric configuration is necessary to fully cancel the phase contribution from constant acceleration \cite{npulse}. An asymmetric configuration could be used to suppress parasitic interferometers, further enhancing contrast \cite{stockton,Dutta}. Although the relative AC Stark shift is small, it  would still result in systematics for the rotation measurement. In order to cancel even this residual relative AC Stark shift, the rotation-sensitive AI is operated with opposite effective wave vectors $+k$ and $-k$.

As shown in Fig. \ref{rotation}(b), the phase sensitivity of the butterfly AI is 40\,mrad for a single fringe measurement, which corresponds to the rotation rate sensitivity of 170\,$\mu$rad/s. By operating the AI on the phase-sensitive slopes of the fringes, we achieve a sensitivity of about 300\,$\mu$rad/s/$\sqrt{\rm Hz}$ for rotation measurement. The Earth's rotation rate is measured by the gyroscope. Using the angled Mach-Zehnder AI, we measure the absolute angle between the laser wave vector and the gravity axis. During the rotation rate measurement, the projection angle is monitored by the tilt sensor. We then obtain the Earth's rotation rate of $5(1)\times 10^{-5}$\,rad/s after 20\,min of averaging time. The expected rate at Berkeley (latitude 37.78$^\circ$) is 57.9\,$\mu$rad/s corresponding to a phase shift of 24\,mrad. The phase sensitivity of our gyroscope is similar to that achieved by the state-of-art atom interferometry gyroscopes \cite{canual,Dutta}. The rotation rate sensitivity is constrained by the total phase accumulated from the rotation, which can be improved by longer sequence time.

\section{Conclusion}

In conclusion, we have demonstrated multiaxis AIs with a single-diode laser system and a pyramidal MOT. Efficient Doppler-sensitive Raman transitions are achieved using a small single photon detuning and modest laser intensity. With zero differential AC Stark shift and insignificant single photon scattering, high-contrast fringes are obtained. Gravity as well as the tilt of the platform are measured in Mach-Zehnder geometry with a sensitivity of 6\,$\mu$m/s$^2/\sqrt{\rm Hz}$ and 4\,$\mu$rad/$\sqrt{\rm Hz}$, respectively. Rotation is measured using the butterfly geometry with a sensitivity of 300\,$\mu$rad/s/$\sqrt{\rm Hz}$.

Being simple, precise, and capable of multiaxis operation, our multiaxis and single-diode AI has the potential to become a versatile atomic sensor in rough environments, such as drones, submarines or satellites. Comparing with classical inertial sensors, AIs are more accurate and have better long-term stability. Generally, AIs are too sensitive to operate in environments with strong vibration noise. In order to overcome this problem, either measuring vibration with another sensor or simultaneous multiple AIs are feasible. One example is that AIs have been operated on an airplane by monitoring the vibration noise with a mechanical accelerometer \cite{Geiger}. A simultaneous dual-species atom accelerometer has been proposed to cancel the common vibration noise \cite{Bonnin}. Extending our multiaxis AI to simultaneous operation will make it immune to vibration and enable more applications. Furthermore, marrying our single-diode atom interferometry with grating-based or tetrahedral MOTs \cite{tetrahedralMOT,gratingMOT,chipMOT,SubDoppler}, more deployable atomic sensors are foreseeable, such as on-chip gravimeters, gradiometers and gyroscopes.

\section*{Funding Information}
Bakar Fellows Program; David and Lucile Packard Foundation; NASA Planetary Instrument Definition and Development Program through a Contract with Jet Propulsion Laboratory.

\section*{Acknowledgments}

We thank Cheong Chan, Tatyana Gavrilchenko, Chen Lai, Weicheng Zhong, and Philipp Haslinger for their contributions to the experiment and discussions. J.D. thanks the NSF Graduate Student Fellowship for support.








\begin{thebibliography}{1}
\bibitem{Cronin} A. D. Cronin, J. Schmiedmayer, and D. E. Pritchard, ``Optics and interferometry with atoms and molecules,'' Rev. Mod. Phys. {\bf 81}, 1051-1129 (2009).

\bibitem{Peters} A. Peters, K. Y. Chung, and S. Chu, ``High-precision gravity measurements using atom interferometry,'' Metrologia {\bf 38}, 25-61 (2001).

\bibitem{GravIso} H. M\"uller, S.-W. Chiow, S. Herrmann, S. Chu, K.-Y. Chung, ``Atom-interferometry tests of the isotropy of post-Newtonian gravity,'' Phys. Rev. Lett. {\bf 100}, 031101 (2008).

\bibitem{Hu} Z.-K. Hu, B.-L. Sun, X.-C. Duan, M.-K. Zhou, L.-L. Chen, S. Zhan, Q.-Z. Zhang, and J. Luo, ``Demonstration of an ultrahigh-sensitivity atom-interferometry absolute gravimeter,'' Phys. Rev. A {\bf 88}, 043610 (2013).

\bibitem{McGuirk} J. M. McGuirk, G. T. Foster, J. B. Fixler, M. J. Snadden, and M. A. Kasevich, ``Sensitive absolute-gravity gradiometry using atom interferometry,'' Phys. Rev. A {\bf 65}, 033608 (2002).

\bibitem{Rosi} G. Rosi, L. Cacciapuoti, F. Sorrentino, M. Menchetti, M. Prevedelli, and G. M. Tino, ``Measurement of the gravity-field curvature by atom interferometry,'' Phys. Rev. Lett. {\bf 114}, 013001 (2015).

\bibitem{Asenbaum} P. Asenbaum, C. Overstreet, T. Kovachy, D. D. Brown, J. M. Hogan, and M. Kasevich, ``Phase shift in an atom interferometer due to spacetime curvature across its wave function,'' Phys. Rev. Lett. {\bf 118}, 183602 (2017).

\bibitem{Gustavson} T. L. Gustavson, P. Bouyer, and M. A. Kasevich, ``Precision rotation measurements with an atom interferometer gyroscope,'' Phys. Rev. Lett. {\bf 78}(11), 2046-2049 (1997).

\bibitem{stockton} J. K. Stockton, K. Takase, and M. A. Kasevich, ``Absolute geodetic rotation measurement using atom interferometry,'' Phys. Rev. Lett. {\bf 107}, 133001 (2011).

\bibitem{Dutta} I. Dutta, D. Savoie, B. Fang, B. Venon, C. L. Garrido Alzar, R. Geiger, and A. Landragin, ``Continuous cold-atom inertial sensor with 1 nrad/sec rotation stability,'' Phys. Rev. Lett. {\bf 116}, 183003 (2016). 

\bibitem{Fixler} J. B. Fixler, G. T. Foster, J.M. McGuirk, and M. A. Kasevich, ``Atom interferometer measurement of the Newtonian constant of gravity,'' Science {\bf315}(5808), 74-77 (2007).

\bibitem{Rosi2014} G. Rosi, F. Sorrentino, L. Cacciapuoti, M. Prevedelli, and G. M. Tino, ``Precision measurement of the Newtonian gravitational constant using cold atoms,'' Nature {\bf 510}, 518-521 (2014).

\bibitem{Lan} S.-Y. Lan, P.-C. Kuan, B. Estey, D. English, J.M. Brown, M. A. Hohensee, and H. M\"uller, ``A clock directly linking time to a particle's mass," Science {\bf 339}(6119), 554-557 (2013).


\bibitem{Bouchendira} R. Bouchendira, P. Clad\'e, S. Guellati-Kh\'elifa, F. Nez, and F. Biraben, ``New determination of the fine structure constant and test of the quantum electrodynamics,'' Phys. Rev. Lett. {\bf 106}, 080801 (2011).


\bibitem{Parker} R. H. Parker, C. Yu, B. Estey, W. Zhong, E. Huang, and H. M\"uller, ``Controlling the multiport nature of Bragg diffraction in atom interferometry,'' Phys. Rev. A {\bf 94}, 053618 (2016).

\bibitem{Zhou} L. Zhou, S. Long, B. Tang, X. Chen, F. Gao, W. Peng, W. Duan, J. Zhong, Z. Xiong, J. Wang, Y. Zhang, and M. Zhan, ``Test of equivalence principle at $10^{-8}$ level by dual-species double-diffraction Raman atom interferometer,'' Phys. Rev. Lett. {\bf 115}, 013004 (2015).

\bibitem{Hartwig} J. Hartwing, S. Abend, C. Schubert, D. Schlippert, H. Ahlers, K. Posso-Trujillo, N. Gaaloul, W. Ertmer, and E. M. Rasel, ``Testing the universality of free fall with rubidium and ytterbium in a very large baseline atom interferometer,'' New J. Phys. {\bf 17}, 035001 (2015).

\bibitem{Duan} X.-C. Duan, X.-B. Deng, M.-K. Zhou, K. Zhang, W.-J. Xu, F. Xiong, Y.-Y. Xu, C.-G. Shaom, J. Luo, and Z.-K. Hu, ``Test of the universality of free fall with atoms in different spin orientations,'' Phys. Rev. Lett. {\bf 117}, 023001 (2016).

\bibitem{Hamilton} P. Hamilton, M. Jaffe, P. Haslinger, Q. Simmons, H. M\"uller, and J. Khoury, ``Atom-interferometry constraints on dark energy,'' Science {\bf 349}(6250), 849-851 (2015).

\bibitem{Jaffe} M. Jaffe, P. Haslinger, V. Xu, P. Hamilton,  A. Upadhye, B. Elder, J. Khoury, and H. M\"uller, ``Testing sub-gravitational forces on atoms from a miniature in-vacuum source mass,'' Nat. Phys. {\bf 13}, 938-942 (2017).

\bibitem{Kovachy} T. Kovachy, P. Asenbaum, C. Overstreet, C. A. Donnelly, S. M. Dickerson, A. Sugarbaker, J. M. Hogan, and M. A. Kasevich, ``Quantum superposition at the half-meter scale,'' Nature {\bf 528}, 530-533 (2015).

\bibitem{Yu} N. Yu and M. Tinto, ``Gravitational wave detection with single-laser atom interferometers,'' Gen. Relativ. Gravit. {\bf 43}, 1943 (2011).

\bibitem{Graham} P. W. Graham, J. M. Hogan, M. A. Kasevich, and S. Rajendran, ``New method for gravitational wave detection with atomic sensors,'' Phys. Rev. Lett. {\bf 110}, 171102 (2013).

\bibitem{Harms} J. Harms, B. J. J. Slagmolen, R. X. Adhikari, M. C. Miller, M. Evans, Y. Chen, H. M\"uller, and M. Ando, ``Low-frequency terrestrial gravitational-wave detectors,'' Phys. Rev. D {\bf 88}, 122003 (2013).

\bibitem{Freier} C. Freier, M. Hauth, V. Schkolnik, B. Leykauf, M. Schilling, H. Wziontek, H.-G. Scherneck, J. M\"uller, and A. Peters, ``Mobile quantum gravity sensor with unprecedented stability,'' J. Phys. Conf. Ser. {\bf 723}, 012050 (2016).

\bibitem{Fang} B. Fang, I. Dutta, P. Gillot, D. Savoie, J. Lautier, B. Cheng, C. L. Garrido Alzar, R. Geiger, S. Merlet, F. Pereira Dos Santos, and A. Landragin, ``Metrology with atom interferometry: Inertial sensors from laboratory to field applications,'' J. Phys. Conf. Ser. {\bf 723}, 012049 (2016).

\bibitem{Geiger} R. Geiger, V. M\'enoret, G. Stern, N. Zahzam, P. Cheinet, B. Battelier, A. Villing, F. Moron, M. Lours, Y. Bidel, A. Bresson, A. Landragin, and P. Bouyer, ``Detecting inertial effects with airborne matter-wave interferometry,'' Nat. Commun. {\bf 2}, 474 (2011). 

\bibitem{Barrett} B. Barrett, L. Antoni-Micollier, L. Chichet, B. Battelier, T. L\'ev\`eque, A. Landragin, and P. Bouyer, ``Dual-matter-wave inertial sensors in weightlessness,'' Nat. Commun. {\bf 7}, 13786 (2016).

%

\bibitem{canual} B. Canuel, F. Leduc, D. Holleville, A. Gauguet, J. Fils, A. Virdis, A. Clairon, N. Dimarcq, C. J. Bord\'e, A. Landragin, and P. Bouyer, ``Six-axis inertial sensor using cold-atom interferometry,'' Phys. Rev. Lett. {\bf 97}, 010402 (2006).

\bibitem{Dickerson} S. M. Dickerson, J. M. Hogan, A. Sugarbaker, D. M. S. Johnson, and M. A. Kasevich, ``Multiaxis inertial sensing with long-time point source atom interferometry,'' Phys. Rev. Lett. {\bf 111}, 083001 (2013).

\bibitem{Hardman} K. S. Hardman, P. J. Everitt, G. D. McDonald, P. Manju, P. B. Wigley, M. A. Sooriyabandara, C. C. N. Kuhn, J. E. Debs, J. D. Close, and N. P. Robins, ``Simultaneous precision gravimetry and magnetic gradiometry with a Bose-Einstein Condensate: a high precision, quantum sensor,'' Phys. Rev. Lett. {\bf 117}, 138501 (2016).


\bibitem{Rakholia} A. V. Rakholia, H. J. McGuinness, and G. W. Biedermann, ``Dual-axis high-data-rate atom interferometer via cold ensemble exchange,'' Phys. Rev. Appl. {\bf 2}, 054012 (2014).



\bibitem{Bodart} Q. Bodart, S. Merlet, N. Malossi, F. Pereira Dos Santos, P. Bouyer, and A. Landragin, ``A cold atom pyramidal gravimeter with a single laser beam,'' Appl. Phys. Lett. {\bf 96}, 134101 (2010).


\bibitem{Butts} D. L. Butts, J. M. Kinast, B. P. Timmons, and R. E. Stoner, ``Light pulse atom interferometry at short interrogation times,'' J. Opt. Soc. Am. B {\bf 28}(3), 416-421 (2011).

\bibitem{Andia}	M. Andia, R. Jannin, F. Nez, F. Biraden, S. Guellati-Kh\'elifa, and P. Clad\'e, ``Compact atomic gravimeter based on a pulsed and accelerated optical lattice,'' Phys. Rev. A {\bf 88}, 031605 (2013).


\bibitem{CavityAI} P. Hamilton, M. Jaffe, J. M. Brown, L. Maisenbacher, B. Estey, and H. M\"uller, ``Atom interferometry in an optical cavity,'' Phys. Rev. Lett. {\bf 114}, 100405 (2015).

\bibitem{Biedermann} G. W. Biedermann, H. J. McGuinness, A. V. Rakholia, Y.-Y. Jan, D. R. Wheeler, J. D. Sterk, and G. R. Burns, ``Atom interferometry in a warm vapor,'' Phys. Rev. Lett. {\bf 118}, 163601 (2017).



\bibitem{Abend}	S. Abend, M. Gebbe, M. Gersemann, H. Ahlers, H. M\"untinga, E. Giese, N. Gaaloul, C. Schubert, C. L\"ammerzahl, W. Ertmer, W. P. Schleich, and E. M. Rasel, ``Atom-chip fountain gravimeter,'' Phys. Rev. Lett. {\bf 117}, 203003 (2016). 

\bibitem{fiberlaser} C. Diboune, N. Zahzam, Y. Bidel, M. Cadoret, and A. Bresson, ``Multi-line fiber laser system for cesium and rubidium atom interferometry,'' Opt. Express {\bf 25}(15), 16898-16906 (2017).

\bibitem{diodelaser} A. N. Dinkelaker, M. Schiemangk, V. Schkolnik, A. Kenyon, K. Lampmann, A. Wenzlawski, P. Windpassinger, O. Hellmig, T. Wendrich, E. M. Rasel, M. Giunta, C. Deutsch, C. K\"urbis, R. Smol, A. Wicht, M. Krutzik, and A. Peters, ``Autonomous frequency stabilization of two extended-cavity diode lasers at the potassium wavelength on a sounding rocket,'' Appl. Opt. {\bf 56}(5), 1388-1396 (2017).

\bibitem{Lee} K. I. Lee, J. A. Kim, H. R. Noh, and W. Jhe, ``Single-beam atom trap in a pyramidal and conical hollow mirror,'' Opt. Lett. {\bf 21}(15), 1177-1179 (1996).

\bibitem{Pollock} S. Pollock, J. P. Cotter, A. Laliotis, F. Ramirez-Martinez, and E. A. Hinds, ``Characteristics of integrated magneto-optical traps for atom chips,'' New J. Phys. {\bf 13}, 043029 (2011).

\bibitem{KasevichChu} M. Kasevich, and S. Chu, ``Atomic interferometry using stimulated Raman transitions,'' Phys. Rev. Lett. {\bf 67}(2), 181-184 (1991).


\bibitem{Biedermann2009} G. W. Biedermann, X. Wu, L. Deslauriers, K. Takase, and M. A. Kasevich, ``Low-noise simultaneous fluorescence detection of two atomic states,'' Opt. Lett. {\bf 34}(3), 347-349 (2009).

\bibitem{Steck} Daniel A. Steck, Cesium D Line Data. Available at http://steck.us/alkalidata/cesiumnumbers.1.6.pdf


\bibitem{ZhouM} M.-K. Zhou, X Xiong, L.-L. Chen, J.-F. Cui, X.-C. Duan, and Z.-K. Hu, ``Note: A three-dimension active vibration isolator for precision atom gravimeters,'' Rev. Sci. Instrum. {\bf 86}, 046108 (2015).


\bibitem{Ahlers} H. Ahlers, H. M\"untinga, A. Wenzlawski, M. Krutzik, G. Tackmann, S. Abend, N. Gaaloul, E. Giese, A. Roura, R. Kuhl, C. L\"ammerzahl, A. Peters, P. Windpassinger, K. Sengstock, W. P. Schleich, W. Ertmer, and E. M. Rasel, ``Double Bragg interferometry,'' Phys. Rev. Lett. {\bf 116}, 173601 (2016).

\bibitem{npulse} M. Cadoret, N. Zahzam, Y. Bidel, C. Diboune, A. Bonnin, F. Th\'eron, and A. Bresson, ``Phase shift formulation for N-light-pulse atom interferometers: application to inertial sensing,'' J. Opt. Soc. Am. B {\bf 33}(8), 1777-1788 (2016).


\bibitem{Bonnin} A. Bonnin, C. Diboune, N. Zahazm, Y. Bidel, M. Cadoret, and A. Bresson, ``New concepts of inertial measurements with multi-species atom interferometry,'' arXiv:1710.06289 (2017).



\bibitem{gratingMOT} M. Vangeleyn, P. F. Griffin, E. Riis, and A. S. Arnold, ``Laser cooling with a single laser beam and a planar diffractor,'' Opt. Lett. {\bf 35}(20), 3453-3455 (2010).

\bibitem{chipMOT} C. Nshii, M. Vangeleyn, J. P. Cotter, P. F. Griffin, E. A. Hinds, C. N. Ironside, P. See, A. G. Sinclair, E. Riis, and A. S. Arnold, ``A
surface-patterned chip as a strong source of ultracold atoms for quantum technologies,'' Nat. Nanotechnol. {\bf 8}, 321–324 (2013).

\bibitem{SubDoppler} J. P. McGilligan, P. F. Griffin, R. Elvin, S. J. Ingleby, E. Riis, and A. S. Arnold, ``Grating chips for quantum technologies,'' Sci. Rep. B {\bf 7}, 834 (2017).

\bibitem{tetrahedralMOT} M. Vangeleyn, P. F. Griffin, E. Riis, and A. S. Arnold, ``Single-laser, one beam, tetrahedral magneto-optical trap,'' Opt. Express {\bf 17}(16), 13601-13608 (2009).

\end{thebibliography}
\end{document}